\documentclass[12pt]{article}

\newcommand{\be}{\begin{equation}}
\newcommand{\ee}{\end{equation}}
\newcommand{\bea}{\begin{eqnarray}}
\newcommand{\eea}{\end{eqnarray}}
\newcommand{\beann}{\begin{eqnarray*}}
\newcommand{\eeann}{\end{eqnarray*}}
\newcommand{\nn}{\nonumber}

\newcommand{\rot}[3]{\left[{#1}\atop{#2}\right]_{#3}}
\newcommand{\p}{\varphi}
\newcommand{\e}{\epsilon}
\newcommand{\diag}{{\rm diag}}
\newcommand{\one}{{\bf 1}}
\newcommand{\Z}{{\bf Z}}
\newcommand{\R}{{\bf R}}
\newcommand{\C}{{\bf C}}
\newcommand{\A}{\mbox{\boldmath$A$}}
\newcommand{\B}{\mbox{\boldmath$B$}}
\newcommand{\x}{\mbox{\boldmath$x$}}
\newcommand{\del}{\partial}
\newcommand{\D}{{\cal D}}
\newcommand{\til}{\widetilde}

\makeatletter
  
  \@addtoreset{equation}{section}
\makeatother

\begin{document}

%
%
\begin{titlepage}

\setcounter{page}{0}
\renewcommand{\thefootnote}{\fnsymbol{footnote}}

\begin{flushright}
KUNS-1702 \\
hep-th/0101082
\end{flushright}

\vspace{15mm}
\begin{center}
{\large\bf
Supersymmetric D-brane Bound States with $B$-field\\
and\\
Higher Dimensional Instantons
on Noncommutative Geometry
}

\vspace{15mm}
{\large
Kazutoshi Ohta\footnote{e-mail address:
kohta@yukawa.kyoto-u.ac.jp}
}\\
\vspace{10mm}
{\em Department of Physics, Kyoto University, Kyoto 606-8502, Japan} \\
\end{center}
\vspace{25mm}
\centerline{{\bf{Abstract}}}
\vspace{5mm}

We classify supersymmetric D0-D$p$ bound states with a non-zero
$B$-field by considering T-dualities of intersecting branes at
angles. Especially, we find that the D0-D8 system with the $B$-field
preserves 1/16, 1/8 and 3/16 of supercharges if the $B$-field
satisfies the ``(anti-)self-dual'' condition in dimension eight. The
D0-branes in this system are described by eight dimensional instantons
on non-commutative $\R^8$. We also discuss the extended ADHM
construction of the eight-dimensional instantons and its deformation
by the $B$-field. The modified ADHM equations admit a sort of the
`fuzzy sphere' (embeddings of $SU(2)$) solution.

\end{titlepage}
\newpage

\renewcommand{\thefootnote}{\arabic{footnote}}
\setcounter{footnote}{0}

\section{Introduction}

D-branes in a background of an NS-NS two form $B$-field have been
attracting much interest in the development of string theory. The
constant magnetic $B$-field on the D-brane in particular gives a
string theoretical realization of the non-commutative geometry
\cite{CDS,DH,SW} and the world-volume effective theory on it is
described by the non-commutative Yang-Mills theory. One of features of
the non-commutative Yang-Mills theory is that a singularity of a
moduli space of instantons on non-commutative geometry is naturally
resolved \cite{NS} and we do not encounter the problems of the
singularity. So the non-commutative geometry is very helpful tool for
considering the resolution of the singularity in the moduli space of
instantons.

Four dimensional $k$ instantons of the gauge group $U(N)$ is realized
as $k$ D0-branes on $N$ D4-branes in Type IIA string theory. If we
turn on the anti-self dual constant $B$-filed, which preserves 1/4 of
supersymmetries, the D0-branes are ``resolved'' and we can not
separate the D0-branes and D4-branes as long as preserving
supersymmetry. If we divert to D0-brane theory of the gauge group
$U(k)$ with $N$ matter multiplets, moduli space of vacua of the Higgs
branch coincides with the moduli space of the instantons and the
$B$-field corresponds to Fayet-Iliopoulos (FI) parameters of the
theory. If the FI parameters are non-zero, the system can not enter
the Coulomb branch which is described by the separation of the D0 and
D4-branes. Therefore, these two pictures are consistent.

On the other hand, these systems are equivalent by string duality to
the other brane configurations which include rotated branes at
angles. The configuration of the intersecting brane at angles has been
applied to the construction of three dimensional Chern-Simons system
on the branes \cite{KOO}. The relation between the non-commutative
nature of the Wilson line operator in Chern-Simons theory and the
non-commutative geometry is discussed in \cite{Kohta2}. So the branes
with the $B$-field gives an alternative understanding of dynamics on
the intersecting brane at angles.

In this way, the D-brane bound states with the $B$-field are very
interesting in the context of both brane dynamics and brane
world-volume theory. Recently, their systems are discussed from
various points of view in \cite{CIMM, OTom, MPT, Witten, BBH, Imamura}. The
authors discuss the condition for the $B$-fields preserving the
supersymmetry. However, some cases of the enhancement of supersymmetry
was missed unfortunately. In the D0-D8 system, there are the cases
preserving 1/16, 1/8 and 3/16 of supercharges, for instance. In this
paper, we classify the suitable supersymmetric configuration of the
constant $B$-field by using the results of the supersymmetric
intersecting brane at angles \cite{S-J, OT}. We find that the $B$-field
must satisfy the self-dual condition in the D0-D4 or D0-D8 system. In
particular, in the D0-D8 system, the $B$-field satisfies the extended
``self-dual'' conditions given by \cite{CDFN, Ward}, which associate
with the symmetry $Spin(7)$, $Spin(6)$ and $Spin(5)$. We obtain only a
part of the components of the ``self-dual'' $B$-field from the
intersecting brane configuration, but the general value of the
$B$-field, which belongs to a representation of the subgroup
$Spin(7)$, $Spin(6)$ and $Spin(5)$ of the eight dimensional rotational
group $SO(8)$, certainly preserves 1/16, 1/8 and 3/16 of
supersymmetries, respectively.

We also discuss in detail the extended ADHM construction of the eight
dimensional ``self-dual'' instantons when the symmetry group is
$Spin(5)\simeq Sp(2)$. If we turn on the constant $B$-field on the
D8-brane, the coordinates gain the non-commutativity. We consider the
extended ADHM equations on the non-commutative geometry provided by the
$B$-field in various representations of $Sp(2)$. We show that the
non-commutativity of the coordinates modifies the ADHM equations and
one of the modified equations admits a `fuzzy sphere' solution, which
is proportional to the generator of the $SU(2)$ algebra. This type of
the solution is a significant feature on the eight dimensional
instantons. This fact means that the moduli space of the eight
dimensional instanton possesses a rich structure and its resolution of
the singularity is more complicated than the four dimensional
case. The application of the non-commutative geometry to the extended
ADHM construction is more useful to understand how the singularity is
resolved.

This paper is organized as follows. In section \ref{M5-brane}, we
revisit the classification of the residual supersymmetry in the
configuration of the intersecting brane at angles. We treat the case
of four angles and increase the number of the angles in sequence. The
fractions 1/16, 1/8, 3/16 and 1/4 of the supersymmetries appear in the
classification.  In next section, we relate the configuration of the
brane at angles with the D0-D$p$ system with the $B$-field. We find
that the supersymmetric configuration of the $B$-field in the D0-D4 or
D0-D8 satisfies the self-dual or extended ``self-dual''
conditions. The fact means that we can expect the general
``self-dual'' $B$-field background on the D8-brane preserves the same
supersymmetry as the same as the case of the D0-D4 with the
anti-self-dual $B$-field.  We use these ``(anti-)self-dual'' $B$-field
configurations and discuss the resolution of the eight dimensional
instantons, which is regarded as the D0-brane on the D8-brane, on the
non-commutative geometry in section \ref{ADHM}. We find that the
``(anti-)self-dual'' $B$-field modifies the extended ADHM
construction of the eight dimensional instantons. We also discuss the
solution of the modified ADHM equations.  The final section is devoted
to discussion and comment.

\section{Intersecting brane at angles and residual supersymmetry}
\label{M5-brane}

We start with a configuration of two intersecting M5-branes at angles
in M-theory and discuss residual supersymmetries. Supersymmetries of
the intersecting M5-branes are completely classified by \cite{S-J, OT}. We
give an account of their work here because of slightly different
notations for the later discussion. They discuss the general rotations
by five angles, but we consider only the case up to four angles since
we would like to obtain the D$p$-D$p'$ system with a $B$-field as we
will discuss in the next section.

We consider the two intersecting M5-branes whose world-volumes are
\bea
{\rm M5} & : & \left( 013579 \right),
\label{M5 config} \\
{\rm M5}' & : & \left( 0\rot{1}{2}{\p_1}\rot{3}{4}{\p_2}\rot{5}{6}{\p_3}\rot{7}{8}{\p_4}9 \right),
\label{M5' config}
\eea
where the numbers in parentheses represent world-volume directions and the symbol $\rot{\mu}{\nu}{\p}$ means that the brane wold-volume is tilted in $(x^\mu,x^\nu)$-plane by an angle $\p$. The presence of these two M5-branes gives a constraint on a 32 components Killing spinor $\e$ in 11 dimensions.
\bea
{\rm M5}  & : & \Gamma_{013579} \e = \e,
\label{M5} \\
{\rm M5}' & : & R \Gamma_{013579} R^{-1} \e = \e,
\label{M5'}
\eea
where
\be
R = \exp \left\{ \pi \sum_{i=1}^{4} \p_i \Gamma_{2i-1, 2i} \right\}
\ee
is a rotation matrix in the spinor representation and $\p_i$ are four angles in the range of $0 \leq \p_i < 1$.
If some of components of this spinor survive after these projections, supersymmetries are unbroken.

We rewrite the eq.~(\ref{M5'}) by using $R\Gamma_{013579}R^{-1}=R^2\Gamma_{013579}$ and eq.~(\ref{M5}) into
\be
(R^2 -1)\e = 0.
\label{proj}
\ee
Introducing the following diagonalized bases of the gamma matrices 
\bea
\Gamma_{013579} & = & \diag\left( \one_{16},-\one_{16} \right),
\label{gammaM5} \\
\Gamma_{1234} & = & \diag\left( \one_{8},-\one_{8},\cdots \right),\\
\Gamma_{1256} & = & \diag\left( \one_{4},-\one_{4},\one_{4},-\one_{4},\cdots \right),\\
\Gamma_{1278} & = & \diag\left( \one_{2},-\one_{2},\one_{2},-\one_{2},\one_{2},-\one_{2},\one_{2},-\one_{2},\cdots \right),
\eea
where dots mean repeating the first 16 diagonal components and $\one_n$ stands for $n \times n$ identity matrix,
then $R^2-1$ in eq.~(\ref{proj}) reduces to
\bea
R^2-1 & = & 2R\Gamma_{12}\nn\\
&& \times \diag\left(
\sin\pi(\p_1-\p_2-\p_3-\p_4) \one_2, \sin\pi(\p_1-\p_2-\p_3+\p_4) \one_2, \right.\nn\\
&& \sin\pi(\p_1-\p_2+\p_3-\p_4) \one_2, \sin\pi(\p_1-\p_2+\p_3+\p_4) \one_2,\nn\\
&& \sin\pi(\p_1+\p_2-\p_3-\p_4) \one_2, \sin\pi(\p_1+\p_2-\p_3+\p_4) \one_2,\nn\\
&& \left. \sin\pi(\p_1+\p_2+\p_3-\p_4) \one_2, \sin\pi(\p_1+\p_2+\p_3+\p_4) \one_2,
\cdots \right).
\label{R2-1}
\eea

When we act this operator on the Killing spinor $\e$, some components
of $\e$ can be non-zero if we choose suitable angles, that is, the
supersymmetry remains. We can count the number of supersymmetries of
the above M5-brane system as the residual spinor components. To see
this, we increase the number of the angles in order setting the others
to be zero. Then we classify the possible number of supersymmetries
and find the condition for the angles.

We notice that the non-rotated M5-brane necessarily breaks half of the
supersymmetries due to the projection of the gamma matrix
(\ref{gammaM5}). Since the projection of the M5-brane preserves only
16 components of the Killing spinor, it is sufficient to consider the
action of (\ref{R2-1}) only on the first 16 components in the
following.

\bigskip
\hspace*{-\parindent}{\it One angle}

If we set $\p_2=\p_3=\p_4=0$, all components in (\ref{R2-1}) do not
vanish unless $\p_1=0$, namely, supersymmetry is completely broken in
general. In the case of $\p_1=0$, the M5-branes represents nothing but
two parallel M5-branes which preserves 1/2 of the supersymmetric
charges.

\bigskip
\hspace*{-\parindent}{\it Two angles}

We set $\p_3=\p_4=0$. Some of elements in (\ref{R2-1}) vanish when
\be
\p_1\pm\p_2 \in \Z.
\label{2angle}
\ee
Here all integer values are not allowed since we restrict the angles
to the range of $0 \leq \p_i < 1$, but we can choose a suitable
integer if possible. This notion applies below.  For the condition of
(\ref{2angle}), the number of elements being zero in (\ref{R2-1}) is 8
of 16, that is, the system with the condition (\ref{2angle}) preserves
1/4 supersymmetry.

\bigskip
\hspace*{-\parindent}{\it Three angles}

We set $\p_4=0$. The condition for preserving supersymmetry is
\be
\p_1\pm\p_2\pm\p_3 \in \Z.
\label{3 angles}
\ee
In this case, 4 of first 16 elements are zero in (\ref{R2-1}) and 4
components in 32 of the Killing spinor survive. So we have 1/8 of
supersymmetries.

\bigskip
\hspace*{-\parindent}{\it Four angles}

In this case, there are the various ways to obtain the supersymmetric
configuration in contrast with the above cases. One is the similar
condition to the other angles, which is
\be
\p_1\pm\p_2\pm\p_3\pm\p_4 \in \Z.
\ee
This preserves 1/16 supersymmetry.

The second is obtained by setting independently
\be
\p_i\pm\p_j \in \Z \quad {\rm and} \quad \p_k\pm\p_l \in \Z.
\label{1/8 angle}
\ee
There are 12 conditions in total.
For example, if we set $\p_1=-\p_2\neq\p_3=-\p_4$, 4 of the first 16 elements in (\ref{R2-1}) vanish. So the supersymmetry is enhanced to 1/8 compared with the previous case.

The final condition is particular. We now set all angles to be the
same up to a suitable choice of relative signs, that is,
\be
\p_1=\pm \p_2 = \pm \p_3 = \pm \p_4.
\ee
If we choose as $\p_1=\p_2=\p_3=\p_4$ for instance, 6 components of the first 16 is zero. So we have 3/16 residual supersymmetries as a result.

\section{Supersymmetric D-brane bound states with a $B$-field}

In this section we produce D$p$-D$p'$ bound states, which preserve
some supersymmetry, from the M5-brane configuration discussed in
section \ref{M5-brane} by using the M-theory compactification on
$T^8\times S^1$.

First we consider the compactification on $S^1$ along the
$x^9$-direction. The M5-brane configuration (\ref{M5 config}) and
(\ref{M5' config}) is dual to the intersecting D4-brane at four angles
in Type IIA theory. We next would like to take T-dualities along
$x^1$, $x^3$, $x^5$ and $x^7$-directions on the torus $T^8$ which
extends along $x^1,\ldots,x^8$. This T-duality simply maps the
M5-brane to the D0-brane, but since the M5$'$-brane is tilted on $T^8$
at some angles, this rotated D4$'$-brane generally produces D8-brane
with a $B$-field on $T^8$ (See for example \cite{Polchinski, CIMM})
and the geometry on $T^8$ will be non-commutative. The $B$-field is
given by
\be
B_{\mu\nu}=\frac{1}{2\pi\alpha'}\left(
\begin{array}{ccccc}
0    & b_1 &        &      &     \\
-b_1 & 0   &        &      &     \\
     &     & \ddots &      &     \\
     &     &        & 0    & b_4 \\
     &     &        & -b_4 & 0
\end{array}
\right),
\ee
where $\mu,\nu=1,\ldots,8$ and $b_i$ are related to the four angles by
\be
\tan2\pi\p_i = -\frac{1}{b_i},
\ee
or equivalently,
\be
e^{4\pi i \p_i}=-\frac{1+ib_i}{1-ib_i}.
\label{angle-b}
\ee

This D0-D8 bound state preserves the same supersymmetry as the
configuration discussed in the previous section since T-duality does
not change the number of the supercharges. Of course, we can consider
from the beginning the conditions for the Type IIA Killing spinors
when the $B$-field exists on the D0-D8 bound state, but it just give
the same conditions as the intersecting brane case. Moreover, there is
no distinction of the twisted boundary condition for the worldsheet
theory between the intersecting brane at angles and D-brane with the
$B$-field. So we can simply apply the supersymmetric classification
for the brane at angles to the D-brane bound state with the $B$-field
by using the translation rule (\ref{angle-b}).

In the classification of the previous section, we set some angles to
be zero in the case that the number of the rotated angles is less than
four. Since this means that some parts of the worldvolume directions
are parallel to M5-brane, T-duality finally makes the D0-D2, D0-D4 and
D0-D6 bound states with the $B$-field, which correspond to the one
angle, two angles and three angles case, respectively.

So we can now classify the supersymmetric condition for the $B$-field
in sequence.

\bigskip
\hspace*{-\parindent}{\it D0-D2 system}

The one angle case corresponds to the D0-D2 bound state with a
constant $B$-field on the D2. For general value of $b_1$, the
supersymmetry is completely broken even if D0-D2 system without the
$B$-field, which corresponds to the case of $\p_1=1/4$ and the others
are zero. An exception is D0-D2 in the limit of $|b_1|\rightarrow
\infty$, but this is just equivalent to the D0-D0 system.

\bigskip
\hspace*{-\parindent}{\it D0-D4 system}

In this case, the supersymmetric condition of the 1/4 supersymmetry
for two angles reduces to
\be
\left(\frac{1+ib_1}{1-ib_1}\right)\left(\frac{1+ib_2}{1-ib_2}\right)^{\pm 1} = 1.
\ee
This means that
\be
B_{12}=\mp B_{34},
\ee
that is, the D0-D4 system is supersymmetric if and only if the $B$-field is anti-self-dual or self-dual.

\bigskip
\hspace*{-\parindent}{\it D0-D6 system}

There are various choices of the 1/8 supersymmetric condition for the
three angles depending on the relative signs of the angles. We here
only consider the case that the signs are all positive without loss of
generality. Then the condition for the $b_i$ is
\be
\prod_{i=1}^{3}\left(\frac{1+ib_i}{1-ib_i}\right)=-1,
\ee
so we find that
\be
B_{12}B_{34}+B_{34}B_{56}+B_{56}B_{12} = 1,\ \cdots.
\ee
It is not clear what this condition means in the sense of the
anti-symmetric two form field.

The D0-D6 bound state without the $B$-field corresponds to the
intersecting branes at angles of $\pm\p_1=\pm\p_2=\pm\p_3=\frac{1}{4}$
and $\p_4=0$. These angles do not satisfy the condition (\ref{3
angles}), so the D0-D6 system can not be supersymmetric if a suitable
$B$-field does not exist.

\bigskip
\hspace*{-\parindent}{\it D0-D8 system}

In this case, there are three patterns depending on the fractions of
the residual supersymmetry. First, we choose one of the conditions as
$\sum_{i=1}^{4}\p_i=0$ which preserves the 1/16 supersymmetry, then
the condition for $b_i$ is
\be
\prod_{i=1}^{4}\left(\frac{1+ib_i}{1-ib_i}\right)=1.
\ee
From this, we find that
\be
\begin{array}{l}
B_{12}+B_{34}+B_{56}+B_{78} \\
\qquad = B_{12}B_{34}B_{56} + B_{34}B_{56}B_{78} + B_{56}B_{78}B_{12} + B_{78}B_{12}B_{34},\ \cdots.
\end{array}
\ee
This condition reduces to the simple form in the small $B$ limit as
\be
B_{12}+B_{34}+B_{56}+B_{78} = 0,\ \cdots.
\label{1/16 B}
\ee
This consists of an extended ``self-dual'' condition in dimensions
greater than four discussed in \cite{CDFN,Ward},
\be
\frac{1}{2}T_{\mu\nu\rho\sigma}B^{\rho\sigma}=\lambda B_{\mu\nu},
\label{SD}
\ee
where $T_{\mu\nu\rho\sigma}$ is a totally antisymmetric tensor and
$\lambda$ is an eigenvalue. The tensor $T_{\mu\nu\rho\sigma}$ is not
invariant under the 8 dimensional rotational group $SO(8)$ but at
least must be invariant under subgroups of $SO(8)$. The above
condition (\ref{1/16 B}) breaks $SO(8)$ to $Spin(7)$ and {\bf 28} of
$B_{\mu\nu}$ decomposes into {\bf 21} which corresponds to $\lambda=1$
and the tensor is given by
\be
T^{\mu\nu\rho\sigma} = \eta^T \gamma^{\mu\nu\rho\sigma} \eta,
\ee
where $\gamma^{\mu\nu\rho\sigma}$ is the totally anti-symmetric
product of $SO(8)$ gamma matrices and $\eta$ is a constant unit
spinor, which satisfies $\eta^T\eta = 1$.

The second condition preserving 1/8 of the supersymmetries is
typically $\p_1 = \p_2 \neq \p_3 = \p_4$. This condition contains two
sets of the self-dual condition on each four dimensional submanifolds
\be
B_{12}=B_{34},\ B_{56}=B_{78},\ \cdots,
\label{1/8 B}
\ee
in terms of the $B$-field. In this case the tensor
$T_{\mu\nu\rho\sigma}$ is a singlet under the subgroup $(SU(4)\times
U(1))/\Z_2$ and (\ref{1/8 B}) is a part of the ``self-dual'' condition
(\ref{SD}) with $\lambda=1$. This type of the condition could be also
considered as an Hermitian-Einstein condition
\be
g^{\alpha\bar{\beta}} B_{\alpha\bar{\beta}} = 0,
\label{HE}
\ee
where $g^{\alpha\bar{\beta}}$ is a hermitian metric on the $T^8$.

Note that the condition (\ref{1/8 B}) or (\ref{HE}) is associated with
the invariance under $SU(4)\simeq Spin(6)$, which is a holonomy of a
Calabi-Yau four-fold (${\rm CY}_4$). This fact means that the
intersecting brane at angles of (\ref{1/8 angle}) or D0-D8 system with
the $B$-field of (\ref{1/8 B}) is dual to the ${\rm CY}_4$.

Finally, we consider the case preserving 3/16 supersymmetries in the
D0-D8 system. One of the conditions is $\p_1=\p_2=\p_3=\p_4$. The
condition is expressed in terms of the $B$-field as
\be
B_{12}=B_{34}=B_{56}=B_{78},\ \cdots.
\label{3/16 B}
\ee
This also is a kind of the extended ``self-dual'' condition which is
invariant under the subgroup $(Sp(2)\times Sp(1))/\Z_2$. Under this
symmetry, the $B_{\mu\nu}$ in the adjoint representation survive as
{\bf 10} when $\lambda=1$ and satisfies the condition (\ref{3/16
B})\footnote{Notice that the choice of the coordinates and the
normalization of $\lambda$ differ from \cite{Ward}.}. The tensor
$T_{\mu\nu\rho\sigma}$ is a singlet of $Sp(2)\simeq Spin(5)$, which is
characteristic symmetry of the D0-D8 system with the B-field of
(\ref{3/16 B}).

In the construction of the $B$-field from the branes at angles, we
have obtained only a part of the components of the ``self-dual''
$B$-field, but we can expect that the general ``self-dual'' $B$-field
preserves the same supersymmetry since the vev of the $B$-field is
compatible with the holonomy of the dual manifold of the D-brane bound
states with the $B$-field. So we treat the components of the
$B$-field as the general ``self-dual'' one in the following.

\section{D0-D8 system with B-field and extended ADHM construction on noncommutative geometry}
\label{ADHM}

\subsection{Noncommutative instanton as D0-D4 with the $B$-field}

The ADHM construction \cite{ADHM, CG} is very important tool for
finding gauge configurations and analyzing moduli space of
instantons. We first briefly review the ADHM construction on four
dimensional space and its modification on the noncommutative
$\R^4$. The noncommutativity of $\R^4$ is realized by introducing a
constant magnetic $B$-field on the D4-brane. The supersymmetric D0-D4
bound state with the $B$-field is regarded as a resolved instanton on
the noncommutative $\R^4$ whose moduli space is non-singular. We in
the following concentrate the case of the gauge group $U(N)$ with
instanton number $k$ since it simply relates to the bound state of $k$
D0-branes and $N$ D$p$-branes.

We first consider the instantons on the ordinary commutative space. In
the ADHM construction, it is useful to treat four coordinates of
$\R^4$ as a quaternion. Introducing the quaternionic basis $\sigma_\mu
= (i\tau_1,i\tau_2,i\tau_3,\one_2)$, where $\tau_i$ are the Pauli
matrices, the coordinates are arrenged into the quaternion variable
\be
\x = \sum_{\mu=1}^{4}\sigma_\mu x^\mu =
\left(
\begin{array}{cc}
z_2 & z_1 \\
-\bar{z}_1 & \bar{z}_2
\end{array}
\right),
\ee
where we define the complex coordinates as $z_1=x^2+ix^1$ and
$z_2=x^4+ix^3$.

We first define an operator 
\be
\D_z = \A + \B\x,
\ee
where $\A$ and $\B$ are $(N+2k)\times 2k$ matrices and the product
with $\x$ is defined in the quaternionic sense as we will see
concretely below.

If we can find a solution of the following equations
\be
\D_z^\dag \psi = 0,
\label{Dpsi}
\ee
for $(N+2k)\times N$ matrix $\psi$, which is normalized as $\psi^\dag
\psi =\one_{N}$, we obtain the $U(N)$ gauge field
\be
A_\mu = \psi^\dag \del_\mu \psi.
\label{gauge field}
\ee

Using a completeness relation
\be
\one_{N+2k} = \psi\psi^\dag + \D_z(\D_z^\dag\D_z)^{-1}\D_z^\dag,
\ee
and assuming that $\D_z^\dag\D_z$ are invertible and commute with all
quaternions, we have the field strength from the gauge field
(\ref{gauge field})
\bea
F_{\mu\nu} & = & 2\psi^\dag\left(\del_{[\mu}\D_z(\D_z^\dag\D_z)^{-1}\del_{\nu]}\D_z^\dag\right)\psi \nn \\
& = & 2\psi^\dag\B \bar{\eta}_{\mu\nu} (\D_z^\dag \D_z)^{-1} \B^\dag \psi,
\label{field strength}
\eea
where
$\bar{\eta}_{\mu\nu}=\frac{1}{2}(\sigma_\mu\sigma_\nu^\dag-\sigma_\nu\sigma_\mu^\dag)$
is a constant self-dual tensor which satisfies
$\frac{1}{2}\e_{\mu\nu\rho\sigma}\bar{\eta}^{\rho\sigma}=\bar{\eta}_{\mu\nu}$. Therefore,
the field strength automatically satisfies the self-dual condition.

The commutativity of $\D_z^\dag\D_z$ with quaternions is the crucial
condition in order to obtain the self-dual field strength. So all of
our tasks finding the self-dual gauge field configuration of
instantons amount to solving the commuting conditions for
$\D_z^\dag\D_z$. From the definition of $\D_z$, we have
\be
\D^\dag_z \D_z = \A^\dag\A + \A^\dag\B\x + \x^\dag\B^\dag\A + \x^\dag\B^\dag\B\x,
\label{D+D}
\ee
then we can rewrite into the condition for the components of $\A$ and
$\B$ with respect to each order in $\x$ since the commuting condition
must satisfy for any $\x$.

Before rewriting the condition, we notice that there are equivalences
between different sets of $\A$ and $\B$ as
\be
\A \sim U \A M, \qquad \B \sim U \B M,
\label{equivalence}
\ee
where $U\in U(N+2k)$ and $M\in GL(2k,\C)$. The gauge field is
invariant under this transformation. Using this symmetry, we can
arrange the matrices $\A$ and $\B$ into
\be
\A = \left(
\begin{array}{cc}
A_2 & A_1 \\
-A_1^\dag & A_2^\dag \\
I & J
\end{array}
\right), \qquad
\B= \left(
\begin{array}{cc}
\one_k & 0 \\
0 & \one_k \\
0 & 0
\end{array}
\right),
\label{diag AB}
\ee
where $A_i$ and $B_i$ are $k\times k$ and $I$ and $J$ are $N\times k$
matrices.  Substituting these matrices into (\ref{D+D}) and requiring
the commutativity of $\D^\dag_z\D_z$ with quaternions, we obtain the
following sets of equations
\bea
\mu_\R & \equiv & [A_2^\dag,A_2] - [A_1^\dag,A_1] + I^\dag I - J^\dag J = 0, 
\label {ADHM R}\\
\mu_\C & \equiv & [A_2^\dag,A_1] + I^\dag J = 0.
\label{ADHM C}
\eea
These real and complex equations are known as the ADHM
equation of instantons. Three degrees of the equations associate with
{\bf 3} of $SU(2)\simeq Sp(1)$, which reflects the hyper-K\"ahler
structure of the instanton moduli space. If we find the solutions of
the ADHM equations, we can construct the self-dual field strength
through $\psi$ which satisfies eq.~(\ref{Dpsi}). Therefore, the moduli
space of instantons is described by the hyper-K\"ahler quotient in the
space of the solutions
\be
{\cal M}= \left(\mu^{-1}_\R (0)\cap \mu^{-1}_{\C} (0)\right)/U(k).
\label{sing moduli}
\ee
This moduli space is singular, but we can resolve the singularity by the modification of the ADHM equations.

Let us next consider the resolution of the ADHM equations by turning
on the $B$-field. In string theory, if the constant $B$-field exists
on the D4-brane the coordinates on the D4-brane become non-commutative
\be
[x^\mu,x^\nu] = i\theta^{\mu\nu}.
\ee
$\theta^{\mu\nu}$ are determined by the $B$-field as
\be
\theta^{\mu\nu} = 2\pi\alpha'
\left(
\frac{1}{g+2\pi\alpha'B}
\right)^{\mu\nu}_{A},
\ee
where the subscript $A$ means an anti-symmetric part of the matrix.

The $B$-field on the D4-brane admits only the self-dual or
anti-self-dual one for preserving supersymmetry. However, the
self-dual $B$-field does not change the ADHM equations of the
self-dual instanton. Therefore, the self-dual $B$-field is useless for
resolving the moduli space (\ref{sing moduli}). However, as we will
see below, a constant anti-self-dual $B$-field deforms the self-dual
ADHM equations.

We first introduce the anti-self-dual $B$-field having only a non-zero
component of
\be
B_{12}=-B_{34}=\beta_1.
\ee
This non-zero $B$-field gives the non-commutativity on $\R^4$ in the complex coordinates
\be
[z_1,\bar{z}_1]=-[z_2,\bar{z}_2]=-\frac{\zeta}{2},
\ee
and others are zero. Here we define $\frac{\zeta}{4} = - \frac{2\pi\alpha'\beta_1}{1+\beta_1^2}$.

From the commuting condition of the operator $\D_z^\dag \D_z$, we find
the modification of the first ADHM equation (\ref{ADHM R})
\be
\mu_\R = \zeta.
\label{ADHM R2}
\ee
The second is not changed.

We can also introduce the following non-zero components of the
$B$-field with keeping the anti-self-dual condition
\be
B_{14}=-B_{23}=\beta_2, \qquad B_{13}=B_{24}=\beta_3.
\ee
This gives
\be
[z_1,\bar{z}_2]=-\rho,
\ee
where $\rho=-2(\beta_2-i\beta_3)/\Delta$ and
$\Delta=1+\beta_1^2+\beta_2^2+\beta_3^2$.  These non-commutative
coordinates modify the condition for the ADHM construction of the
instantons and we have in addition to (\ref{ADHM R2})
\be
\mu_\C=\rho.
\ee

As a result, the singularity of the moduli space of instantons on the
commutative $\R^4$ is resolved to a smooth manifold
\be
\tilde{{\cal M}}= \left(\mu^{-1}_\R (\zeta)\cap \mu^{-1}_{\C} (\rho)\right)/U(k).
\ee

Here we see the constant anti-self-dual $B$-field modifies the ADHM
equations of the self-dual instantons. If we would like to consider
the ADHM construction of the anti-self-dual instantons, one may
exchange a role of $\sigma_\mu$ and $\sigma_\mu^\dag$. Then, we have a
anti-self-dual tensor
$\eta_{\mu\nu}=\frac{1}{2}(\sigma_\mu^\dag\sigma_\nu-\sigma_\nu^\dag\sigma_\mu)$
in the formula (\ref{field strength}) instead of
$\bar{\eta}_{\mu\nu}$. The self-dual $B$-field only affects the ADHM
equations of the anti-self-dual instantons vice-versa.


\subsection{Resolution of D0 in D8 by the B-field}

We now consider the extended ADHM construction of the eight
dimensional ``self-dual'' instantons with $Sp(2)$ symmetry given in
\cite{CGK}, and its modification by the $B$-field.

In order to treat the eight dimensional space, we first provide two
quaternionic coordinates in which we arrange the eight coordinates as
\be
\x = \sum_{\mu=1}^{8} \til{\sigma}_\mu x^{\mu} =
\left(
\begin{array}{cc}
z_2 & z_1 \\
-\bar{z}_1 & \bar{z}_2
\end{array}
\right),\quad
\x' = \sum_{\mu=1}^{8} \til{\sigma}'_\mu x^{\mu} =
\left(
\begin{array}{cc}
z_4 & z_3 \\
-\bar{z}_3 & \bar{z}_4
\end{array}
\right),
\ee
where we define the eight vector matrices
\bea
\til{\sigma}_\mu & = & (i\tau_1,0,i\tau_2,0,i\tau_3,0,\one_2,0), \nn \\
\til{\sigma}'_\mu & = & (0,i\tau_1,0,i\tau_2,0,i\tau_3,0,\one_2), \nn
\eea
and the four complex coordinates $z_1=x^3+ix^1$, $z_2=x^7+ix^5$,
$z_3=x^4+ix^2$ and $z_4=x^8+ix^6$.

The Dirac-like operator in the ADHM construction extends to
\be
\D_z = \A + \vec{\B}\cdot\vec{\x},
\ee
where $\vec{\B}=(\B,\B')$ and $\vec{\x}=(\x,\x')$. The matrices $\A$, $\B$ and $\B'$ are $(N+2k)\times 2k$ similar to the previous case.

The construction of instantons is also similar to the four dimensional
case. Finding the solution of the equation (\ref{Dpsi}) at first and
using the relation
\be
\Sigma_\mu\equiv \del_\mu \vec{\x} = 
\left(
\begin{array}{cc}
\til{\sigma}_\mu \\
\til{\sigma}'_\mu
\end{array}
\right),
\ee
then we find the ``self-dual'' field strength
\bea
F_{\mu\nu} & = & 2\psi^\dag\left(\del_{[\mu}\D_z(\D_z^\dag\D_z)^{-1}\del_{\nu]}\D_z^\dag\right)\psi \nn \\
& = & 2\psi^\dag\vec{\B} \bar{N}_{\mu\nu} (\D_z^\dag \D_z)^{-1} \vec{\B}^\dag \psi,
\eea
where
$\bar{N}_{\mu\nu}=\frac{1}{2}(\Sigma_\mu\Sigma_\nu^\dag-\Sigma_\nu\Sigma_\mu^\dag)$
is a ``self-dual'' tensor which satisfies
$\frac{1}{2}T_{\mu\nu\rho\sigma}\bar{N}^{\rho\sigma}=\lambda\bar{N}_{\mu\nu}$
with $\lambda=1$.

Here we require the condition that $\D_z^\dag \D_z$ should commute with
$\Sigma_\mu$. This is a necessary condition to obtain the
``self-dual'' gauge configurations on the eight dimensions.

In this construction, there are equivalences between different sets of
$\A$, $\B$ and $\B'$ as like as (\ref{equivalence}). Using this
symmetry, we can rearranged $\A$ and $\B$ as in (\ref{diag AB}) and
$\B'$ as\footnote{Precisely speaking, the degrees of freedom of the
relative coordinate choices in $\x$ and $\x'$ still remain, but we
have fixed them for later convenience.}
\be
\B' = \left(
\begin{array}{cc}
B_2 & B_1  \\
-B_1^\dag & B_2^\dag \\
K & L
\end{array}
\right).
\ee
In this representation of matrices, the commuting condition of
$\D_z^\dag \D_z$ gives the following sets of equations
\bea
\mu_\R^1 & \equiv & [A_2^\dag,A_2] - [A_1^\dag,A_1] + I^\dag I - J^\dag J = 0,
\label{8 ADHM R1} \\
\mu_\C^1 & \equiv & [A_2^\dag,A_1] + I^\dag J = 0, \\
\mu_\C^2 & \equiv & [A_2^\dag,B_2] - [B_1^\dag,A_1] + I^\dag K - L^\dag J = 0, \\
{\mu_\C^2}' & \equiv & [A_2^\dag,B_1] + [B_2^\dag,A_1] + I^\dag L + K^\dag J = 0, \\
\mu_\R^3 & \equiv & [B_2^\dag,B_2] - [B_1^\dag,B_1] + K^\dag K - L^\dag L = 0, \\
\mu_\C^3 & \equiv & [B_2^\dag,B_1] + K^\dag L = 0.
\eea
There are two real and four complex equations, which relate to the
adjoint representation {\bf 10} of $Sp(2)$. So we expect that the
moduli space of the eight dimensional instantons considering now has a
structure of $Sp(2)$ holonomy. An example of the manifold with $Sp(2)$
holonomy is known as the toric hyper-K\"ahler manifold \cite{GGPT} and
a relation between the 't Hooft type solution of the above extended
ADHM equations and the manifold with $Sp(2)$ holonomy is discussed in
\cite{PT}.

We next consider the effect of the $B$-field on the ADHM construction
in eight dimensions. In our considering case, the tensor
$T^{\mu\nu\rho\sigma}$ in the ``(anti-)self-dual'' condition is
invariant under the subgroup $(Sp(1)\times Sp(2))/\Z_2$ of
$SO(8)$. Under this symmetry, {\bf 28} of anti-symmetric two form $B$
in eight dimensions decomposes into {\bf 10}, {\bf 3} and {\bf 15},
which satisfy the ``(anti-)self-dual'' condition (\ref{SD}) with
$\lambda=1, -\frac{5}{3}, -\frac{1}{3}$, respectively. Therefore,
there are one ``self-dual'' and two ``anti-self-dual'' $B$-field,
which are compatible with $Sp(2)$ symmetry, and we may have the
modification of the extended ADHM equations by these $B$-fields.

First, we consider the case that the $B$-field is {\bf 10} with
$\lambda=1$. We obtain the following commutation relations of the
complex coordinates from the non-zero components of the $B$-field
\bea
&& [z_1,\bar{z}_1]=[z_2,\bar{z}_2]=-\frac{\zeta_1}{2}, \qquad [z_3,\bar{z}_3]=[z_4,\bar{z}_4]=-\frac{\zeta_2}{2},\nn \\
&& [z_1,z_2]=-\rho_1, \quad [z_1,\bar{z}_3]=[z_2,\bar{z}_4]=-\rho_2,\nn \\
&& [z_1,z_4]=-[z_2,z_3]=-\rho_3, \quad [z_3,z_4]=-\rho_4,\nn
\eea
and the others are zero, where $\zeta_i \in \R$ and $\rho_i \in \C$
are parameters which are determined by the components of the
$B$-field. The extended ADHM equations in this background are modified
while the $B$-field is ``self-dual'' in contrast with the four
dimensional case. In fact, if we require the commuting condition of
$\D_z^\dag\D_z$, the first equation (\ref{8 ADHM R1}) becomes
\be
\mu_\R^1 = -\bar{\rho}_2 B_2 + \rho_2 B_2^\dag - \bar{\rho}_3 B_1 + \rho_3 B_1^\dag,
\label{mod 8 ADHM R1}
\ee
that is, linear terms in $B$ add to the quadratic ADHM
equation. However, unfortunately, the solution of the above equation
is not so interesting. The l.h.s and r.h.s of (\ref{mod 8 ADHM R1})
must be independently zero since the l.h.s is hermitian but r.h.s is
not. If we choose as $B_2=\rho_2 H_2$ and $B_1=\rho_3 H_1$, where
$H_i$ are hermitian matrices, the equation (\ref{mod 8 ADHM R1}) is
satisfied. Therefore, the moduli space of the eight dimensional
instantons almost is not modified and we can not avoid the
singularity. The ``self-dual'' $B$-field is not useful for the
resolution of the ``self-dual'' instanton moduli space, consequently.

The next case is {\bf 3} with $\lambda=-\frac{5}{3}$. If we define
$N_{\mu\nu}=\frac{1}{2}(\Sigma_\mu^\dag\Sigma_\nu-\Sigma_\nu^\dag\Sigma_\mu)$,
this satisfies the ``anti-self-dual'' condition with
$\lambda=-\frac{5}{3}$. From the three independent components of the
$B$-field, we find the non-commutativity of the complex coordinates
\bea
&& [z_1,\bar{z}_1]=-[z_2,\bar{z}_2]=[z_3,\bar{z}_3]=-[z_4,\bar{z}_4]=-\frac{\zeta}{2}, \nn \\
&& [z_1,\bar{z}_2]=[z_3,\bar{z}_4]=\rho, \nn
\eea
and the other commutation relations are zero. These commutation
relations modify the extended ADHM equations as
\bea
\mu_\R^1 & = & \zeta \left(\one_k+ \Xi \right),\\
\mu_\C^1 & = & \rho \left(\one_k+ \Xi \right),
\eea
where $\Xi = \frac{1}{2}\left( \{B_2^\dag,B_2\} + \{B_1^\dag,B_1\} + K^\dag K + L^\dag L \right)$. This is very similar to the four dimensional instanton except for the existence of $\Xi$. In particular, the first two equations are the same as the resolved ordinary ADHM equation when $B_1=B_2=0$ and $K=L=0$, namely the moduli space of the eight dimensional instanton includes the resolved four dimensional instanton moduli space.

Finally, we consider most interesting case which has a rich
structure. The $B$-field of {\bf 15} obeys the ``anti-self-dual''
condition with $\lambda=-\frac{1}{3}$ and has 15 independent
components. The 15 components are arranged into one real parameter
$\zeta$ and seven complex parameters $\rho_i$ and we have the
following commutation relations
\bea
&& [z_1,\bar{z}_1]=-[z_2,\bar{z}_2]=-[z_3,\bar{z}_3]=[z_4,\bar{z}_4]=-\frac{\zeta}{2}, \nn \\
&& [z_1,\bar{z}_2]=-[z_3,\bar{z}_4]=\rho_1, \quad [z_1,\bar{z}_3]=-[z_2,\bar{z}_4]=-\rho_2, \nn \\
&& [z_1,z_4]=[z_2,z_3]=\rho_3, \quad [z_2,\bar{z}_3]=-\rho_4, \quad [z_1,\bar{z}_4]=\rho_5,\nn \\
&& [z_2,z_4]=\rho_6, \quad [z_1,z_3]=\rho_7.\nn
\eea
These commutation relations gives the modification of the ADHM equations
\bea
\mu_\R^1 & = & \zeta (\one_k- \Xi ) + \bar{\rho}_2 B_2 + \rho_2 B_2^\dag - \bar{\rho}_3 B_1 - \rho_3 B_1^\dag,\\
\mu_\C^1 & = & \rho_1 (\one_k- \Xi ) - \bar{\rho}_4 B_2 + \rho_5 B_2^\dag + \bar{\rho}_6 B_1 - \rho_7 B_1^\dag.
\eea
In this case we have linear terms in $B_i$ again. However, in contrast
with the previous case, there are non-trivial solutions proportional
to $k$ dimensional representation of $SU(2)$ algebra, $T_\pm = T_1\pm
i T_2$ and $T_3$. Actually, if we tune the parameters to
$\zeta=\rho_1=0$ and
$\rho_2=\rho_3=2\rho_4=2\rho_5=2\rho_6=2\rho_7=m$, then we find a
solution
\bea
&& A_2=m T_-, \quad A_1=m T_-, \quad
 B_2=m T_3, \quad B_1= m T_3, \nn \\
&& I=m\lambda_1, \quad J=m \lambda_2, \quad K=m \lambda_3, \quad L=m \lambda_4, \nn
\eea
where $\lambda_i$ are $N \times k$ matrices which satisfy that
$\lambda_1^\dag\lambda_4=\lambda_2^\dag\lambda_3=T_3$,
$\lambda_1^\dag\lambda_1=\lambda_2^\dag\lambda_2$,
$\lambda_3^\dag\lambda_3=\lambda_4^\dag\lambda_4$ and the other
$\lambda_i^\dag\lambda_j\ (i\neq j)$ are zero. For example, when $N=4$
and $k=2$ if we choose $\lambda_1=-i\Sigma_5$, $\lambda_2=-i\Sigma_6$,
$\lambda_3=\Sigma_8$ and $\lambda_4=\Sigma_7$, all equations are
satisfied. The existence of these solutions means that the moduli
space of the eight dimensional instantons is divided into disconnected
pieces every representation of $SU(2)$.

From a point of view of the D0-brane world-volume theory, the matrices
$A_i$ and $B_i$ correspond to the adjoint vector multiplets of the
gauge group $U(k)$, and $I,J,K,L$ are $N$ matter multiplets. The ADHM
equations discussed above are flat conditions which describe the
moduli space of vacua in D0-brane world-volume theory. If we add the
$B$-field background, the flat conditions are changed. One of the
modifications appears as the FI terms as like as the ordinary
supersymmetric Yang-Mills theory of D0-D4 system. However, the linear
term in $B_i$ means the mass term in the system. So the $B$-field also
provides mass of the vector multiplets. Similar modifications of the
flat direction appear in the context of supersymmetric
Chern-Simons theory \cite{Dunne,KL,Kao,Kohta1}, the dielectric effect
of the branes \cite{Myers} and the ${\cal N}{=}1^*$ theory
\cite{PS}. Presumably, the brane realization of these systems may
relate to the D0-D8 with the $B$-field by the string dualities.

\section{Discussion and Comment}

In this paper, we discussed the supersymmetric configuration of the
D0-D$p$ system with the $B$-field. We obtain the supersymmetric system
by using the T-dualities from the intersecting branes at four
angles. In the classification of the supersymmetric intersecting
M5-branes at angles \cite{OT}, we can maximally rotate the brane by up
to five angles. The rotation by five angles includes 1/32, 3/32 and
5/32 supersymmetric cases in addition to what we have
discussed. However, if we naively apply the dimensional reduction and
the T-dualities to these configurations in M-theory, we obtain a bound
state of D0 and $p$ D8-brane with a $B$-field and $q$ rotated
NS5-branes at angles, where $p$ and $q$ are co-prime integers determined
by the fifth angle. This system also preservers the same
supersymmetries as the intersecting branes, but the fractions of the
supersymmetry sound strange in a sense of the D-brane bound states
with the $B$-field. To understand this system we may need more
knowledge about the brane dynamics near the NS5-brane \cite{EGKRS}.

The brane systems with the $B$-field is considered as a dual to the
other brane configurations. The low energy effective theories on these
branes are described by non-commutative Yang-Mills theory,
Chern-Simons theory, ordinary Yang-Mills theory with non-zero FI parameters,
etc. The dynamics of these theories are closely related with each other
by dualities in string theory. The careful analyses of the D-brane
bound states with the $B$-field and their duals will shed light on the
problems of various field theories.

\section*{Acknowledgments}

I would like to thank T.~Yokono and Y.~Michishita for useful discussions and comments. This work is supported in part by JSPS Research Fellowships for Young Scientists.

%
%
\newpage

\newcommand{\NP}[3]{Nucl.~Phys.~{\bf #1} (#2) #3}
\newcommand{\PL}[3]{Phys.~Lett.~{\bf #1} (#2) #3}
\newcommand{\PR}[3]{Phys.~Rev.~{\bf #1} (#2) #3}
\newcommand{\PRL}[3]{Phys.~Rev.~Lett.~{\bf #1} (#2) #3}
\newcommand{\AP}[3]{Ann.~Phys.~{\bf #1} (#2) #3}
\newcommand{\CMP}[3]{Comm.~Math.~Phys.~{\bf #1} (#2) #3}
\newcommand{\JHEP}[3]{JHEP {\bf #1} (#2) #3}
\newcommand{\hepth}[1]{{\tt hep-th/#1}}

\newcommand{\lit}[3]{#1, ``#2'', #3.}

\end{document}